\documentclass[twocolumn]{article}

\usepackage{epsfig}

\input xy
\xyoption{all}
\usepackage{amssymb,amsmath,dsfont}

\newtheorem{definition}{Definition}
\newtheorem{theorem}{Theorem}
\newtheorem{proposition}{Proposition}
\newtheorem{lemma}{Lemma}
\newtheorem{corollary}{Corollary}

\title{Quotient Manifold Projections and Hierarchical Dynamics}

\author{Martin Nilsson Jacobi \\
		Complex Systems Group\\
	         Department of Energy and Environmental Research  \\
		Chalmers University of Technology, 412 96 Gothenburg, Sweden \\
		Ph: +46 (0) 31 772 3166\\
		{\tt mjacobi@chalmers.se}}

\begin{document}
\maketitle
\abstract{
In this paper we explore the mathematical structure of hierarchical organization in smooth dynamical systems. We start by making precise what we mean by a level in a hierarchy, and how the higher levels need to respect the dynamics on the lower levels. We derive a mathematical construction for identifying distinct levels in a hierarchical dynamics. The construction is expressed through a quotient manifold of the phase space and a Lie group that fulfills certain requirement with respect to the flow. We show that projections up to higher levels can be related to symmetries of the dynamical system. We also discuss how the quotient manifold projections relate to invariant manifolds, invariants of the motion, and Noether's theorem. }






%
\section{Introduction}
\label{intro}
In this paper we look at dynamical systems that allow (or do not allow) multiple simultaneous levels  of description. The conceptual framework for this was outlined in a previous publication, see~\cite{hier,hier_alife}. From a technical perspective, the first issue is to  define levels in the hierarchy. With a definition in place we can  seek  tools that can be used to identify the levels, as well as maps between levels. We argue that a reasonable definition of hierarchical levels is to require the dynamics to be self-contained, which in the current presentation means deterministic, at each level within a hierarchy.  A higher level of description is defined through a projective map of the degrees of freedom on a lower level, and the induced map of the dynamics.  Whether, or not, a projective map constitutes a new level of description can be determined by studying the induced dynamics on the higher level, i.e., by  determining whether the map induces a well defined flow describing a new deterministic dynamics with fewer degrees of freedom.

\begin{figure}
\centerline{\psfig{file=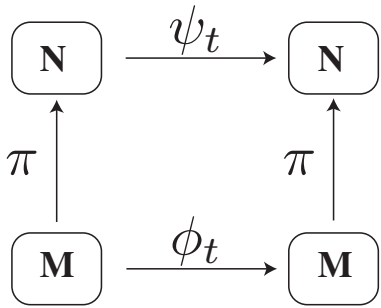,width=5cm}}
\caption{The manifold $M$ is the original phase space, $N$ is a lower dimensional phase space,
and $\phi _t$ and  $\psi _t$ denotes flows on $M$ respective $N$. The projective map $\pi$ describes a new level of description if the diagram commutes. On a more conceptual level, the diagram can be viewed as a definition of (a restricted type of) emergence. 
}
\label{commuting_diagram}
\end{figure}

This can be formulated in the language of differential geometry. Assume that, on the lowest level, the dynamics of the system is described by a trajectory in the phase space space $M$, where $M$ is a manifold. The dynamics, or the trajectory, is an integral curve of a vector field, the infinitesimal generator of a flow $\psi _t$. A transition to a higher level of description, i.e., a dimensional reduction, is described by a projective map\footnote{By projective map we mean a map from one manifold to another manifold with lower dimensionality, possibly a submanifold of the previous} $\pi$ from $M$ to a lower dimensional manifold $N$. The central question is whether, or not, there exists a well defined dynamics, a flow $\psi _t$, on $N$ that describe the projected trajectory. We can express this as a commuting diagram, see Fig.~\ref{commuting_diagram}.
It should be made clear that the approach to hierarchical dynamics, or elimination of degrees of freedom, taken in this paper is purely geometric. This means that we consider only eliminations though projections on the phase space manifold. The projection cannot, for example, depend on time or on the vector field generating the dynamics. To make this distinction clear, consider a periodic orbit of a dynamical system. One possible viewpoint would be to state that the dynamics is one-dimensional. The trajectory can be mapped onto the circle by using the arc length as a parametrization. This projection is however not geometric, and therefore not generally acceptable in the framework considered in this paper. For more details on the periodic orbit see Section~\ref{periodic_orbits}.

\section{Background} 
Reducing the effective dimensionality of a dynamical system is of interest in almost all fields of science that use mathematical modeling. The actual methods for dimensional reduction that are used in different areas vary quite dramatically, both technically but also with respect to what is meant by a dimensionally reduced model. Here I briefly outline three different approaches.
\subsection{Noether's theorem in classical mechanics}
\label{classical_mech}
The problem of eliminating degrees of freedom in dynamical systems has its longest historical legacy in classical mechanics, dating back to Poincare's classic study of many body problem as well as Jacobi's elimination of nodes, see Arnold~\cite{arnold} and references therein (especially~\cite{smale}, as well as~\cite{marmo}). The most elegant and universal formulation of the process was formulated by Noether in her famous theorem relating fundamental variational symmetries with (local) conservation laws, see e.g.,~\cite{goldstein}. For finite dimensional systems, a conservation law is equivalent to an invariant of the motion, which in turn implies that the dynamic occurs on an invariant constraint manifold. Elimination of degrees of freedom can then, at least in principle, be carried out by expressing the dynamics in a coordinate system where the constraint manifold is trivial, i.e., where the constants of the motion separate from the "effective" degrees of freedom. A classic example is the $n$-body problem with conservative pairwise central force interaction. Naively this system has $6 \cdot N$ degrees of freedom. Spatial invariance (translation and rotation) and time translation invariance lead to conservation of momentum, angular momentum and total energy. The total degrees of freedom in the system is therefore reduced by, at least, $3 + 3 + 1 = 7$. 
\subsection{Inertial manifolds}
\label{inv_man}
Central to the study of nonlinear dynamical systems is the identification and computation of invariant manifolds~\cite{Wiggins,Guckenheimer}, or, for infinite dimensional dissipative driven (parabolic)  partial differential equations, inertial manifolds~\cite{Foias}.  The idea is that the dynamics of the system, after a short transient, approaches a positively invariant manifold, or an attractor\cite{Ruelle2}, of significantly lower dimensionality than that of the full system. The effective dynamics of the system can then be described in terms of a parametrization of the invariant manifold, i.e., a low dimensional representation of the system. This scenario is often referred to as slaving of the fast degrees of freedom to the slowly varying degrees of freedom spanning the invariant manifold. Important for this idea to apply is that there exist a clear separation of time scales. Intuitively we can argue that the fast contracting directions in the phase space are effectively removed from the dynamics and the remaining degrees of freedom is the attractor. For a recent overview over this approach to dimensional reduction, see~\cite{model_reduction,gorban}. The general idea of slaving fast degrees of freedom and adiabatic elimination is also a recurring theme in Haken's work on self-organization~\cite{haken}.
\subsection{Projection operators in non-equilibrium statistical physics}
In non-equilibrium statistical physics one usually performs model reduction by focusing attention on physically interesting variables. The affect of the other, uninteresting, degrees of freedom is included as a noise in an effective stochastic differential equation, i.e., a Langevin equation involving the relevant degrees of freedom. Similarly to the situation with invariant manifolds, separation of time scales is important for the noise to be uncorrelated in time (white) and for the higher level dynamics to lack memory, i.e., be "Markovian"~\cite{zwanzig}. In contrast to the situation with invariant manifolds, there are no analytic tools for choosing the relevant degrees of freedom. In practice the choices are guided by physical intuition. When the relevant variables are chosen, however, there are formal techniques for deriving the higher level dynamics. One such method is the use of projection techniques, see the book by Zwanzig~\cite{zwanzig} for a detailed review on this. The classic book by Gardiner\cite{gardiner}  is also a good reference on general procedures for model  reduction in stochastic differential equations. Of special interest in this area is the recent methods for approximating fast chaotic (mixing) degrees of freedom as white noise through an adiabatic elimination procedure, see e.g.,~\cite{just}. For a review of both inertial manifolds and stochastic approximations, see~\cite{givon}.
\subsection{Decomposable dynamics}
\label{decompose}
The techniques discussed in Section~\ref{classical_mech} and~\ref{inv_man} are based on the idea that the dynamics is constrained to a low dimensional manifold, which is embedded in the full phase space. In contrast, the Langevin-type of dynamics, used for dimensional reduction in non-equilibrium statistical mechanics, is derived under the assumption that we know the variables that we are interested in and therefore can treat the rest of the system either as noise (the fast degrees of freedom) or as an external slaving parameters (the slow degrees of freedom). The hierarchical dynamics that we are interested in here is different from both these approaches. We do not assume that the dynamics has an actual dimensionality that is lower than indicated by the naive phase space description. The existence of constraint, or inertial, manifolds is, as we shall see, only a special case. Nor do we assume that we have any a priori knowledge of which degrees of freedom to focus on for the reduction. The aim is to derive the possible projections that reduce the dimensionality of a given dynamical system. Finally, a separation of time scales is in general not assumed.

To demonstrate these points let us consider a simple situation where we have two independent dynamical systems, say the Lorenz system and the R\"{o}ssler system, both with parameter values in the chaotic regimes. Assume further that we are not presented with these system in their "nicest form," i.e., a decomposition where the degrees of freedom separate into two non-interacting $3$-dimensional subgroups. Instead the dynamics is described in terms of some arbitrary nonlinear combination where all degrees of freedom are coupled. Finally, we assume that the two systems' time scales are tuned so the recurrence times on the respective attractor are comparable, i.e., there is no obvious separation of time scales between the Lorenz system and the R\"{o}ssler system. In this situation, none of the aforementioned techniques provides any hope of dimensional reduction, except perhaps to a dimension given by the addition of the respective attractors in the underlying systems. The total system is truly $6$-dimensional. Furthermore, treating some degrees of freedom as noise makes no sense. The technique presented in this paper is however able to decompose the total system into its two "irreducible" parts. As outlined in Section~\ref{intro}, a projection that eliminates the Lorenz part of the dynamics, results in a deterministic dynamics on the higher level, namely the R\"{o}ssler dynamics, and vice versa. There are two different possible projections for the system under consideration. The current framework  focuses on decomposition of dynamical systems. The situation can however be asymmetric. Consider a situation where subsystem $A$ affects subsystem $B$, but not vice versa. In this case, a projection that eliminates subsystem $B$ and results in the dynamics of $A$ is allowed, since $A$ by itself is a well defined dynamical system. Elimination of subsystem $A$ is, however, not allowed since $B$ is not an autonomous dynamical system. A more technical discussion on these issues is given in Section~\ref{inv_man_inv_of_mot}.
\section{Projection Techniques}
\label{ProjTech}
We assume that a transition to a higher level can be described by a (sufficiently) smooth map  $\pi : M \rightarrow N$, where $N$ is the target manifold, and $\mbox{rank} ( \pi ) = \dim  (N) = n < m$ is usually assumed to be constant on $M$. The  rank-deficiency guarantees a decrease of the degrees of freedom. It is of course trivial to write down projective maps $\pi : M \rightarrow N$, since any smooth rank deficient function  works. However, a non-trivial restriction on $\pi$  enters if we require that the map should produce a well defined dynamics on $N$. By "well defined" we mean that the system's time evolution can be described in terms of the coarse grained variables alone, i.e., the dynamics on the higher level is deterministic. Technically we note that since $\pi$ is a smooth  map from $M$ to $N$, it induces a differential map between the respective tangent bundles $\pi _* : \left. TM \right| _x \rightarrow  \left. TN \right| _{\pi (x)}$. Since $\pi$ is not a diffeomorphism, $\pi _*$ does not generally define a new vector field on $N$. There is no guarantee that $\pi (x) = \pi (y)$ implies $\pi _* ( {\bf v} |_x ) = \pi _* ( {\bf v} |_y )$. This means that the induced dynamics on $N$ is usually not well defined (not deterministic). We are interested  in the cases when $\pi$ actually defines a new deterministic dynamical system on a higher level. Therefore we make the following definition:
\begin{definition}
Let $\pi : M \rightarrow N$ be a smooth map from a manifold $M$, with dimension $m$, to a manifold $N$, with dimension $n$, where $n<m$. We assume that ${\bf v}$ is a vector field on $M$. If the differential, $\pi _* : \left. TM \right| _x \rightarrow \left.  TN \right| _{\pi(x)}$ maps ${\bf v}$ onto a well defined vector field ${\bf w}$ on $N$, i.e.,  $\left. {\bf w} \right| _{\pi (x)} = \pi _* ( \left. {\bf v} \right| _x )$ for all $x$ in $M$, then we call $\pi$ a projective fiber map with respect to ${\bf v}$. Furthermore, we define the dynamical system $\left( N , {\bf w} \right)$ to be a new (higher) level of description derived from $\left( M , {\bf v} \right)$.
\label{projective_fiber_maps}
\end{definition}
where we let $\left( M , {\bf v} \right)$ denote a dynamical systems in terms of a phase space $M$ and a generating vector field ${\bf v}$.

The conclusion is that a system allows a higher level of descriptions, according to our definition, if and only if there exists a projective fiber map with respect to the flow (or, to be more exact,  the vector field generating the flow). Note also that the infinitesimal generators of the dynamics ${\bf v}$ and ${\bf w}$ are $\pi$-related, see~\cite{olver} for details.

Assume that $\pi$ is a projective fiber map and that the two points $x$ and $y$ on $M$ are mapped to the same point on $N$, i.e., $\pi (x) = \pi (y)$. The trajectories on $M$ passing though $x$ and $y$ must then map onto the same trajectory on $N$, otherwise the dynamics on the higher level is not deterministic. We conclude that $\pi ( \exp (t {\bf v} x ) = \pi ( \exp (t {\bf v} y )$ either for all $t$, or for no $t$. We use this observation to prove the following lemma   providing conditions on projective fiber maps:
\begin{lemma}
\label{conditions}
Let $\pi : M \rightarrow N$ be a smooth map, and ${\bf v}$ be a vector field on $M$. Assume further that the range of $\pi$ is all of $N$. If $\pi _* ( \left. {\bf v} \right| _x ) = \pi _* ( \left. {\bf v} \right| _y )$ whenever $\pi (x) = \pi (y)$, for all $x$ and $y$ in $M$, then $\pi$ is a projective fiber map with respect to ${\bf v}$. Furthermore, $\pi$ maps integral curves of ${\bf v}$ on integral curves of $\pi _* ( {\bf v} )$, i.e.,
\begin{eqnarray}
  \pi \circ \exp \left( t {\bf v} \right) & = & \exp \left( t \pi _* ( {\bf v} )  \right) \circ \pi ,  \;\;\;\;  \forall t . 
\label{proj_com}
\end{eqnarray}
\end{lemma}
{\bf Proof}:  Assume that 
\begin{eqnarray}
	\pi ( \exp ( t {\bf v} ) x ) & = & \pi ( \exp ( t {\bf v} ) y )
\label{traj}
\end{eqnarray}
for $t=0$, i.e., $\pi (x) = \pi (y)$. We want to show that $\pi _* ( \left. {\bf v} \right| _x ) = \pi _* ( \left. {\bf v} \right| _y )$ then becomes a necessary and sufficient condition for $\pi$ to be a projective fiber map. A Taylor expansion of Eq.\ref{traj} gives
\begin{eqnarray}
	\pi (x) + t {\bf v} ( \pi ) |_x + \dots & = & \pi (y) + t {\bf v} ( \pi )  |_y + \dots .
\label{taylor}
\end{eqnarray}
The higher order terms are irrelevant since successive infinitesimal moves on the trajectory can be used as initial points for a new expansion to first order. Eq.~\ref{taylor} implies ${\bf v} \pi |_x = {\bf v} \pi |_y$, and since ${\bf v} \pi |_x = \pi _* ( {\bf v} |_x )$ this proves Lemma~\ref{conditions}.
$\Box$

Note that Eq.~\ref{proj_com} agrees with the diagram in Fig.~\ref{commuting_diagram}. Note also that the proof is very similar to the proof of Prop.~\ref{symm_func} on symmetries for algebraic equations. This is hardly surprising since $\pi$ is an algebraic function on $M$.
\subsection{Quotient manifold projection}
A projective fiber map has a non-trivial kernel. Concentrating on the kernel, there is a natural approach for constructing projective fiber maps with respect to a given vector field.  The general idea is to define a Lie group action on the manifold $M$, and then let the sub-manifold spanned by the group action define the kernel of the projection $\pi$. The target manifold $N$ then becomes a quotient manifold, i.e., $\pi : M \rightarrow M/G$, and the dimensionality of $N$ is $n = m-k$, where $\dim (M) = m$ and $\dim (G) =k$. A similar approach to dimensional reduction can be found in Landi {\em et. al.}~\cite{landi}. There, however, the analysis is restricted to Hamiltonian systems. Recall the main properties of quotient manifold construction:
\begin{figure}[t]
\centerline{\psfig{file=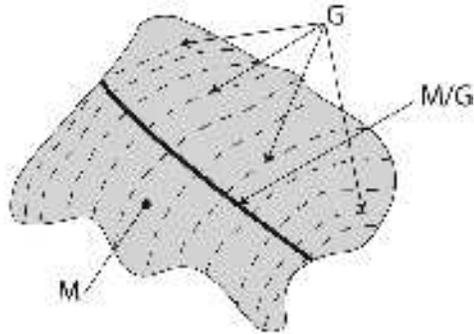,width=6.5cm}}
\caption{The figure shows a sketch of the quotient manifold construction. The gray area  symbolize the manifold $M$. The dashed lines represent the orbits of the Lie group $G$, i.e., the fibration. The thick line is the resulting quotient manifold $M/G$.}
\label{quotient_man_fig}
\end{figure}
\begin{theorem}
Let $M$ be a smooth $n$-dimensional manifold. Suppose $G$ is a local group of transformations which acts regularly on $M$ with $s$-dimensional orbits (through the natural action $\Psi : G \times M \rightarrow M$). Then  there exist a smooth $(n-s)$-dimensional manifold, called the quotient manifold of $M$ by $G$ and denoted $M/G$, together with a projection $\pi : M \rightarrow M/G$, which satisfies
the following properties.
\begin{enumerate}
\item The projection $\pi$ is a smooth map between the manifolds.
\item The points $x$ and $y$ lie on the same orbit in $M$ if and
only if $\pi (x) = \pi (y)$.
\item If ${\mathfrak g}$ denotes the Lie algebra of infinitesimal
generators of the action of $G$, then the linear map
$\pi _* : \left. T M \right| _x  \rightarrow  \left. T \left(
M/G \right) \right| _{\pi (x)}$
is onto, with $\ker \left( \pi _* \right) = \left. {\mathfrak g} \right| _x
= \left\{ \left. {\bf w} \right| _x : {\bf w} \in {\mathfrak g} \right\}$.
\end{enumerate}
\label{quotient_man}
\end{theorem}
The conditions in Lemma~\ref{conditions} need to be transformed into  conditions on the Lie group $G$. The global transformations are hard to address directly. However, the infinitesimal generators of the group, i.e., the Lie algebra, provide all (local) information about the group. The conditions for the quotient manifold projection to be a projective fiber map, is therefore formulated in terms  of the Lie algebra ${\mathfrak g}$ that generates the Lie group $G$, or more precisely the vector fields spanning ${\mathfrak g}$.
\begin{figure}[t]
\centerline{\psfig{file=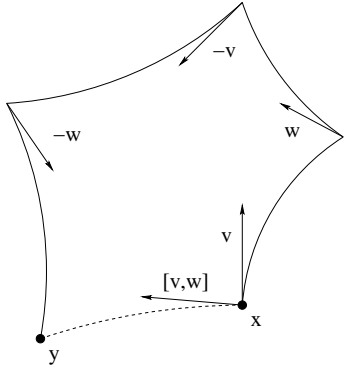,width=4.5cm}}
\caption{A geometric picture of the Lie bracket.}
\label{lie_bracket}
\end{figure}
Let the vector fields spanning ${\mathfrak g}$ be denoted as  $\left\{ {\bf w} _i \right\} _{i=1} ^k$. From the geometric picture of the Lie bracket (see Fig.~\ref{lie_bracket}), we make the following intuitive argument: Let one of the vectors in the Lie bracket be the generator of the flow, and let the other be one  of the generators of ${\mathfrak g}$. When the quotient manifold is constructed, the orbits generated by ${\mathfrak g}$  shrink to a singular point (the point $x$ in the diagram). For this to be a well defined operation, the end points in the diagram in Fig~\ref{lie_bracket} must coincide:  $\left[ {\bf w}, {\bf v} \right] \in {\mathfrak g}$. This becomes our condition for the quotient manifold projection to define a projective fiber map.
\begin{proposition}
\label{quotient_proj}
Let ${\bf v}$ be a vector field on a manifold $M$. Let ${\mathfrak g}$ be the Lie algebra ${\mathfrak g}$ generating a Lie group $G$, with a natural action on $M$. Then the natural projection associated with the quotient manifold $\pi : M \rightarrow M/G$ is a projective fiber map if  $\left[ {\bf v}, {\mathfrak g} \right] \subseteq {\mathfrak g}$ everywhere on $M$. 
\end{proposition}
The notation $\left[ {\bf v}, {\mathfrak g} \right]$ means the set of vectors fields formed by taking members of ${\mathfrak g}$ and commuting them with ${\bf v}$. 

{\bf Proof:} Let $\Psi : G \times M \rightarrow M$ denote a transformation group describing the action of $G$ on $M$, i.e., the natural action of $G$ on $M$. Note that $\pi (x) = \pi ( y )$ then implies $y = \Psi (g, x)$ for some $g \in G$, and the projection fulfills
\begin{eqnarray*}
  \pi \circ \Psi _g & = & \pi \;\;\;\;\;\;\; \forall g \in G .
\end{eqnarray*}
Expressed in terms of the Lie algebra the action of $\Psi$ can be written as:
\begin{eqnarray*}
         y & = & \exp \left( \epsilon  {\bf w} \right) x  ,
\end{eqnarray*}
for some ${\bf w} \in {\mathfrak g}$ and $\epsilon \in {\mathbb R}$.

Further, we use the Campbell-Baker-Hausdorff formula\cite{BCH} on the  form
\begin{eqnarray*}
	\exp ({\bf v}) \exp ({\bf w}) & = & \exp ( \widetilde{\bf w} ) \exp ( {\bf w} ) 
	\exp ( {\bf v} ) , 
\end{eqnarray*}
where 
\begin{eqnarray*}
\widetilde{\bf w} & = & \left[ {\bf v} , {\bf w} \right] + \frac{1}{2} \left(
\left[ {\bf w} , \left[ {\bf w} , {\bf v} \right] \right] - \left[ {\bf v} , \left[ {\bf v} , 
{\bf w} \right] \right] \right) + \cdots .
\end{eqnarray*}
All the terms in the expansion involves deeper nested Lie brackets. From the condition $\left[ {\bf v}, {\mathfrak g} \right] \subseteq {\mathfrak g}$ and the vector space property of a Lie algebra, it is then clear that 
$\widetilde{\bf w} \in {\mathfrak g}$.

We then have
\begin{eqnarray*}
	\pi _* [ \left. {\bf v} \right| _y ] & = &
	\left. \frac{d}{d \tau } \right| _{\tau = 0} 
	\pi \left[ \exp ( \tau {\bf v} ) \exp ( \epsilon {\bf w} )  x  \right] \\
	& = & \left. \frac{d}{d \tau} \right| _{\tau = 0} 
	\pi \left[ \exp ( \widetilde{\epsilon} \; \widetilde{ {\bf w} }) 
	  \exp ( \epsilon  {\bf w} )  \exp ( \tau {\bf v}  ) x \right] \\
	& = & \left. \frac{d}{d \tau} \right| _{\tau = 0}
	\left( \pi \circ \Psi _g \right) \left[ \exp ( \tau {\bf v} ) x
	 \right] \\
	& = & \left. \frac{d}{d \tau} \right| _{\tau = 0}
	 \pi  \left[ \exp ( \tau {\bf v} ) x \right]\\
	& = & \pi _* \left[ \left. {\bf v} \right| _x \right]
\end{eqnarray*}
for some $g \in G$. By this we have shown that $\pi (x) = \pi (y)$ implies that $\pi _* ( \left. {\bf v} \right| _x ) = \pi _* ( \left. {\bf v} \right| _y )$ for all $x$ and $y$ in $M$. $\Box$
\subsection{Constructing $\pi$ from $G$}
\label{constructing_pi}
We now describe how to derive the projective fiber map corresponding to a set of vector fields that fulfill the conditions in  Prop.~\ref{quotient_proj} . By definition $\pi ( \Psi (g , x ) ) = \pi (x)$ for every element $g$ in $G$. The function $\pi$ is therefore $G$-invariant, which implies
\begin{eqnarray}
	{\bf w} _i ( \pi ) & = & 0 ,
\label{const_pi}
\end{eqnarray}
for all ${\bf w} _i \in {\mathfrak g}$. Eq.~\ref{const_pi} is usually derived  by taking the derivative of $\pi ( \exp ( \epsilon {\bf w} ) x ) = \pi (x)$, with respect to $\epsilon$ at $\epsilon = 0$.

In local coordinates, if ${\bf w} _{i} = \sum _{\beta =1} ^n  \eta ^{\beta} _i (x) \frac{\partial}{\partial x^{\beta}}$ the constraint  can be written as a set of quasi-linear first order partial differential equations:
\begin{eqnarray}
	\sum _{\beta =1} ^n \eta _{i} ^{\beta} (x) \frac{\partial \pi ^{\alpha} }{\partial x^{\beta}} 
	& = & 0 ,
\label{constr_pi_direct}
\end{eqnarray}
for $i = 1 , \dots, k$ and $\alpha = 1 , \dots , n$. To find an explicit expression for $\pi$ we need to recursively solve this system,  using e.g., the method of characteristics. 

The analysis shows that the projection $\pi$ consists of invariants of the orbits on $M$ generated by $G$. It should be noted that the generator of the flow, ${\bf v}$, trivially fulfills the requirements in Prop.~\ref{quotient_proj}, since $[ {\bf v} , {\bf v} ] = 0$. If we let $G$ be generated only by ${\bf v}$, however, the dynamics on $N$ is trivial since the trajectory of the system collapses to a singular point. In this special case, all the components of $\pi$ define invariants of the flow, first integrals of the system that is. The existence of such a $\pi$ essentially constitute the starting point of the study of integrable systems. We know that far from  all systems are integrable, i.e., there need not exist $n-1$ invariants of the motion. This brings attention to another important point. The partial differential equations defined in Eq.~\ref{const_pi} do not necessarily have global solutions. Local solutions near non-degenerate points may not always be smoothly extended to global solutions. Determining conditions for when they do is indeed a non-trivial task, and is beyond the current presentation. The ${\mathfrak g}$ spanned by ${\bf v}$ construction also demonstrates another important technical difficulty in the construction of quotient manifolds. Consider, for the sake of the argument, a chaotic system with a strange attractor, such as the famous Lorenz system. An attempt at constructing a new manifold by the quotient $E^3/G$, where $G$ is generated by ${\bf v}$ and $E^3$ is the three dimensional Euclidian space, i.e., the original  phase place, results in a bizarre manifold. The main problem is that the topology of the quotient does not fulfill the Hausdorff property. This is clear from the fractal geometry of the strange attractor. This observation is general, a quotient manifold is not guaranteed to be a Hausdorff space. Some more discussions along these lines follow in Section~\ref{inv_man_inv_of_mot}.
\subsection{Projective fiber maps and symmetries}
A special example of a Lie algebra that fulfills the requirement in Prop.~\ref{quotient_proj} is when
\begin{eqnarray}
	[ {\bf v} , {\bf w} _i ] & = & 0 
\label{symmetry}
\end{eqnarray}
for all ${\bf w} _i \in {\mathfrak g}$. This condition actually implies that ${\bf w}$ is an infinitesimal generator to a symmetry group of the flow generated by ${\bf v}$, see Appendix~\ref{symmetries_ODE} for details. We have the the following general result:
\begin{proposition}
\label{sym_proj}
Let ${\bf v}$ be a vector field on a manifold $M$. Let ${\bf w} _i$ be generators of a symmetry group $G$ of ${\bf v}$, i.e., $[ {\bf v}, {\bf w} _i ] = 0$. Any $G$-invariant function $\pi$, i.e., any function $\pi$ such that ${\bf w} _i ( \pi ) = 0$ $\forall i$, is then a projective fiber map.
\end{proposition}
It may seem like the projective maps found through symmetry fields only form a subset of all projective fiber maps covered by Prop.~\ref{quotient_proj}. However the author suspects that this may not be the case. For a linear dynamics it will be shown later in this paper that all projective maps are indeed generated by symmetries. 
An argument for why this should generalize to the nonlinear case is currently work in progress.

\subsection{Invariant and inertial manifolds, invariants of the motion, integrability, and Noether's theorem}
\label{inv_man_inv_of_mot}
In this section we revisit the discussion started in Section~\ref{decompose}. We can now understand the relation between invariant manifolds, invariants of the motion, and projective fiber  maps in more detail. We start by noting that the existence of invariants of the motion is directly related to the existence of an invariant manifold. To see this, assume that the invariant manifold is defined as an  implicit sub-manifold, i.e., as the solution surface to some smooth constraint function $F (x) = C$, for some constant $C$ that may depend on the initial value of the trajectory. By definition then, $F(x)$ is an invariant of the motion since the trajectory is bound to stay on the  invariant manifold.

How is this related to projective fiber maps? A dynamical system with $n$ degrees of freedom and $k$ invariants of the motion $F _i (x)$, $i = 1, \dots , k$, can be transformed into a form where the constraint surface is separated from the dynamics. This is achieved by making a change of variables where $k$ degrees of freedom are set to $y_i = F_i (x)$ $i = 1 , \dots, k$, and the rest of the degrees of freedom are given by arbitrary functions $y_ {k+j} = G _j (x)$, $j = 1 , \dots, n - k$. The only constraint is that the functions $F_i$ and $G_j$ are all functionally independent. In this situation, a map that eliminates degrees of freedom corresponding to the constants $y _i$, or any subset of $y _i$ for that matter, is a projective fiber map. This follows since the dynamics of $y _{k+j}$ is well defined. On the other hand, a projection that eliminates $y _{k+j}$ is also a projective fiber map since the dynamics of $y_i$ is trivial, in fact constant $\dot{y} _i =0$, and therefore well defined. From this argument it is clear that elimination of degrees of freedom through invariant manifolds, or invariants of the motion, are special cases of projective fiber maps, namely cases where there exist projective fiber maps onto higher level systems with constant dynamics $\dot{y} = 0$. In this context it is also worth mentioning that invariants of the motion are found through projective fiber maps exactly when ${\bf v} \in {\mathfrak g}$, or, to put it differently, when the ${\bf v} ( f )$ has non-trivial solutions. In the special case when ${\bf v} ( f )$ has $n-1$ functionally independent solutions, i.e., there exist $k=n-1$ invariants of the motion, the dynamical system is often referred to as integrable, which means that it can be reduced to a system that can be solvable by quadrature.

Naively, it may seem like the situation when the eliminated subsystem has a trivial dynamic is somehow degenerate. The situation is however more general than it seems at first. Consider the rather general situation when the eliminated dynamical subsystem has a fixed point. Corresponding to that fixed point the total system has an invariant manifold, defined by trajectories starting with a subset of the degrees of freedom at the fixed point of the subsystem. Eliminating the subsystem with the fixed point, e.g., through a projective fiber map,  means projecting the dynamic onto the corresponding invariant manifold. The stability of the invariant manifold is given by the stability of the fixed point. Furthermore, if the eliminated subsystem has many fixed points, each corresponds to an invariant manifold. Projection onto any of these manifolds results in equivalent dynamics on the higher level.

Another important point to discuss is how Corollary~\ref{sym_proj} may be viewed as a generalization of Noether's theorem, see e.g.,~\cite{goldstein} . Recall that Noether's theorem deals with conservative systems in classical mechanics, or, to be more precise, with system whose trajectory is defined as a stationary point of an action integral, i.e., Lagrange variation principle. For such systems, Noether's theorem  ensures that, {\em  for every variational symmetry there exists an invariant of the motion}. For a detailed definition of variational symmetries we refer to Olver~\cite{olver}. Key to understanding the how Corollary~\ref{sym_proj} generalizes Noether's theorem is to note that in the former the focus is on symmetries of the equations of motion\footnote{In the variational framework, the equations of motion are the Euler-Lagrange equations.} which is not the same as variational symmetries considered in the latter. The generalization follows from the fact that every symmetry of the equation of motion is also a variational symmetry, but not vice versa. In other words, there are more possible symmetries that can generate projective fiber maps than variational symmetries that lead to invariants of the motion. From the previous paragraph we may be a bit more precise:
\begin{proposition}
\label{noether}
Consider a system whose trajectories are defined by variation of a Lagrangian $L$. A variational symmetry of $L$ then corresponds to a projective fiber map $\pi$ with a trivial higher level dynamics, i.e., $\pi _* ( {\bf v} |_x ) = 0$ for all $x$.

Conversely, let $\pi$ be a projective fiber map with a non-trivial dynamic on the higher level, then the corresponding symmetry of the equations of motion is not a variational symmetry.
\end{proposition}
The simplest example to illustrate Prop.~\ref{noether} is perhaps two uncoupled harmonic oscillators, see also Section~\ref{linear} and~\ref{periodic_orbits}.

\begin{figure}[t]
\centerline{\psfig{file=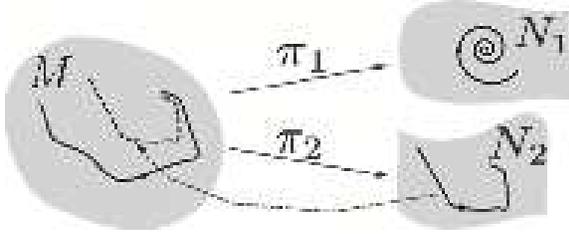,width=7.5cm}}
\caption{The figure is an illustration of the decomposition of a dynamical system on a manifold $M$. In this case we imagine that the dynamics revealed  through the projection $\pi _1$ has a (globally) stable fixed point. In this case the dynamics on $M$ is attracted to a submanifold corresponding to the dynamics of the subsystem given by $\pi _2$ on $N _2$. This is a special case on the inertial manifold picture. In a more general situation, the projective fiber map $\pi _1$ would  {\em not exist} since the location of the fixed point would change as a function of the location on $N _2$, i.e., the subsystem on $N _1$ would be "slaved" to the dynamics of the subsystem on $N_2$. If the dynamics on $N_1$ is fast enough to always relax to the fixed point, the system is still constrained to a submanifold. This is the general picture of an inertial manifold.}
\label{Decomp_subjugation}
\end{figure}

Another interesting situation is when the dynamics converge quickly to a submanifold, an inertial manifold, of lower dimensionality than the total phase space. In this case there are usually no exact invariants of the motion. Still the dynamics of the system has fewer effective degrees of freedom than naively expected. As discussed e.g., in~\cite{Foias,model_reduction,gorban} the existence of inertial manifolds can be proven for a wide class of dynamical systems, especially in hyperbolic differential equations, or, to be more conceptual, in driven dissipative systems. Fig.~\ref{Decomp_subjugation} illustrates how inertial manifolds fit into the framework of projective fiber maps.
\section{Illustrative examples}
To gain intuition for the results in the previous section we study three simple, yet non-trivial, examples. We start with a linear system.
 \subsection{Linear dynamics}
\label{linear}
Start by considering a linear dynamical system and a projective map $\pi$ of the form:
\begin{eqnarray}
	\dot{x} & = & A x \nonumber \\
	\pi (x) & = & P x ,
\label{linear_system}
\end{eqnarray}
where $A$ is an $m \times m$ matrix and $P$ is an $n \times m$ matrix, $n <m$. The phase spaces are Euclidean: $M = {\mathbb R} ^m$ and  $N = {\mathbb R} ^n$. 
\begin{proposition}
\label{linear_case}
Consider a linear dynamical system and a linear projective map, as defined in Eq.~\ref{linear_system}. Then $P$ describes a projective fiber map if and only if $\ker ( P )$ in $A$-invariant.\footnote{By invariance under matrix multiplication we always mean left multiplication, i.e., $x \in \ker ( P )$ implies that $A x \in \ker ( P )$}
\end{proposition}
{\bf Proof} : For pedagogical reasons it is instructive to prove the proposition using two different routes. First we simply start by using Lemma~\ref{conditions}, i.e., we show that $P x = P y$ implies $\pi _* ( {\bf v} | _x ) = \pi _* ( {\bf v} | _y )$. Later we show that Prop.~\ref{quotient_proj} can be used to arrive at Prop.~\ref{linear_case} more quickly. 

{\bf Route A:} Start by noting that linearity implies
\begin{eqnarray*}
	P x = P y	& \Leftrightarrow & x - y \in \ker ( P ) .
\end{eqnarray*} 
In the linear case, there is a direct equivalence between the base manifold $M$ and the tangent space $T M | _x$, which simplifies the differential map\footnote{Remember that the differential is locally defined by the Jacobian, and in the linear case the local and global differential are equivalent.} $\pi _* ( {\bf v} | _x ) = P A x$, therefore

\begin{eqnarray*}
	\pi _* ( {\bf v} | _x ) = \pi _* ( {\bf v} | _y ) & \Leftrightarrow &
		A ( x - y) \in \ker ( P ) .
\end{eqnarray*}
It follows that $\ker ( P )$ is $A$-invariant.

Conversly, assume that $\ker (P)$ is not $A$-invariant, i.e., there exists a $z \in \ker (P)$ such that $A z \notin \ker (P)$. Take an arbitrary  point $z ^{\prime}$ and construct $z ^{\prime \prime} = z ^{\prime} - z$, so that $z  = z ^{\prime} - z ^{\prime \prime}$. Now $z \in \ker (P)$ implies $P z ^{\prime} = P z ^{\prime \prime}$, but, since $P A (  z ^{\prime} - z ^{\prime \prime} ) \neq 0$, $P A z ^{\prime} \neq P A z ^{\prime \prime}$. Then,   according to Lemma~\ref{conditions}, $P$ cannot be a projective fiber map. $\Box$

{\bf Route B:} Alternatively, we can use Prop.~\ref{quotient_proj}. Let the vector fields spanning ${\mathfrak g}$ be ${\bf w}_i = \sum _{\alpha} w_{i} ^{\alpha}   \frac{\partial}{\partial x^{\alpha}}$. According to Theorem~\ref{quotient_man} and Eq.~\ref{const_pi} we have 
\begin{eqnarray*}
	\ker (P) & = & \mbox{span} \left( {\bf w}_1 , \dots , {\bf w}_k \right ) .
\end{eqnarray*}
Since the condition $[ {\bf v}, {\mathfrak g} ] \subseteq {\mathfrak g}$ simplifies to
\begin{eqnarray}
	A {\bf w}_i  & = & \sum _j k_{i j} {\bf w} _j ,
\label{lin_com}
\end{eqnarray}
for some structure constants $k_{i j}$, it is clear that $\ker (P)$ is $A$-invariant. $\Box$
\subsection{The circle}
\label{periodic_orbits}
We now analyze the simplest dynamical system that allows a projective fiber map onto a non-trivial manifold, namely 
the circle trajectory corresponding to the linear ordinary differential equation $\dot{x} = A x$ given explicitly by
\begin{eqnarray*}
	\dot{x} & = & -y \\
	 \dot{y} & = & x .
\end{eqnarray*}
The system is linear, so naively we might expect the analysis in the previous section to be sufficient. However, the eigenvalues are purely imaginary $\pm  i$, and the only linear projective fiber map that can be constructed from the eigenvalues is $\pi (x) = P x = ( A - i  ) ( A + i ) x = 0$. However, by extending our searching to more general projections than linear, we can find non-trivial projective fiber maps. In practice we know that the system is best analyzed in a cylindrical coordinate system, but here we ignore this and just apply the machinery "blindly." Let a possible vector field ${\bf w} \in {\mathfrak g}$ be explicitly expressed as 
\begin{eqnarray*}
 	{\bf  w} & = & \eta _x ( x , y ) \frac{\partial}{\partial x} + \eta _y ( x , y ) \frac{\partial}{\partial y} .
\end{eqnarray*} 
Then
\begin{eqnarray*}
	[ {\bf v} , {\bf w} ] & = & \left( -  y \frac{\partial \eta _x}{\partial x} + x \frac{ \partial \eta _x}{\partial y} + \eta _y \right) \frac{\partial }{\partial x} \\ 
	&& + \left( - y \frac{\partial \eta _y}{\partial x} + x \frac{ \partial \eta _y}{\partial y} - \eta _x \right) \frac{\partial }{\partial y} .
\end{eqnarray*}
The condition $[ {\bf v} , {\bf w} ]Ê= 0$ gives  the two relations
\begin{eqnarray*}
	- y \frac{\partial \eta _x}{\partial x} + x \frac{ \partial \eta _x}{\partial y} + \eta _y & = & 0 \\
	- y \frac{\partial \eta _y}{\partial x} + x \frac{ \partial \eta _y}{\partial y} - \eta _x & = & 0 .
\end{eqnarray*}
It is easy to find a linear solution to these equations:
\begin{eqnarray*}
	\eta _x (x,y) & = & A x + B y \\
	\eta _y (x,y) & = &  - B x + A y .
\end{eqnarray*}
It remains to solve Eq.~\ref{const_pi}:
\begin{eqnarray*}
	(A x + B y) \frac{\partial \pi (x,y) }{\partial x} + (- B x + A y ) \frac{\partial \pi (x,y) }{\partial y} & = & 0 .
\end{eqnarray*}
Solutions exist whenever at least one of the parameters is zero: $A=1$, $B=0$, with the solution
\begin{eqnarray*}
	\pi (x,y) & = & F_1 \left( x/y \right) ,
\end{eqnarray*}
for $y \neq 0$, and
\begin{eqnarray*}
	\pi (x,y) & = & F _2 \left( y/x \right) ,
\end{eqnarray*}
for $x \neq 0$, and some arbitrary functions $F_1$ and $F_2$ such that $F_1 (z) = F_2 (z)$ in the domain where both functions are defined;  and $A=0$, $B=1$, with the solution
\begin{eqnarray*}
	\pi (x,y) & = & G \left( x^2 + y^2 \right) ,
\end{eqnarray*}
for some arbitrary function $G$. The last parameter configuration corresponds to ${\bf w} =  {\bf v}$ and the projection is onto an invariant of the motion corresponding to an invariant manifold.  The higher level of description is in this case trivial, with no time dynamics. Further, it is clear that ${\bf v}$, i.e., rotation, is a variational symmetry of the system and therefore Noether's theorem gives us the conserved quantity $G (x^2 + y^2)$. 

The first case is best understood if we chose the functions $F_1 ( x/y ) = \arctan (x/y)$ and $F_1 ( x/y ) = \pi / 2 - \arctan (y/x)$, which results in a parametrization of the circle $S ^1$ by the rotation angle. 

Note that the projective fiber map is defined by two overlapping "coordinate charts:" $\pi (x,y) = F_1 (x/y)$ when $y \neq 0$, and $\pi (x,y) = F_2 (y/x)$ when $x \neq 0$. The situation with different local coordinate maps is generic. The global projective fiber map is constructed from the overlapping local charts, a standard procedure for constructing maps between manifolds.  This means that the target manifold, $N$ can have a nontrivial topological structure, as in this example when the original manifold $M$ was ${\mathbb R}^2$ but the projected dynamics takes place on $N = S ^1$. The freedom to chose arbitrary functions $F$ and $G$ reflects the freedom of a diffeomorphic change of variables on the target manifold.
\subsection{General projection of linear dynamics onto the real projective plane}
\label{gen_linj}
In general, any linear ODE $\dot{x} = A x$ has two trivial symmetries: ${\bf w} _1 = \sum _{i j} A_{i j} x_i \frac{\partial}{\partial x_j}$ and  ${\bf w} _2 = \sum _i  x_i \frac{\partial}{\partial x_i}$. The first symmetry is just the dynamics itself and the corresponding projection maps onto an invariant of the motion. The latter symmetry comes from the trivial observation that the identity matrix commutes with $A$, but it actually gives a non-trivial projective fiber map. In the case of the circle discussed above it resulted in a projection onto the rotation angle, i.e., $S^1$. In the general case the projection must fulfill
\begin{eqnarray*}
	 \sum _j  x_j \frac{ \partial \pi _i (x) }{\partial x_j} & = & 0 ,
\end{eqnarray*}
for all components $i$. The general solution reads
\begin{eqnarray*}
	\pi _i (x) & = &  F_i \left(  \frac{x_i}{x _j} \right) .
\end{eqnarray*} 
For some arbitrary $j$. Note that $\pi _j$ is constant. This reflects the reduction of dimensionality by the projective map. Just as in the case of the circle there is no single projective map valid over the entire phase space. The projective map provides a "coordinate charts" valid in regions where $x_j \neq 0$. The resulting manifold can in fact be identified as the real projective plane, $P {\mathbb R} ^{n-1}$, if the original dynamics was in ${\mathbb R}^n$. Note that $P {\mathbb R}^1 \simeq S^1$.
\subsection{Subjugated degrees of freedom}
Consider the dynamical system with the following nonlinear skew-product structure:
\begin{eqnarray*}
	\dot{x} & = & f(x) \\
	\dot{y} & = & g(x,y) .
\end{eqnarray*}
It is clear that a projection onto the first degree of freedom, $\pi (x,y ) = F(x)$ for some arbitrary function $F$, is a projective fiber map. It is, however, still interesting to see how this reflects in the formalism. Let ${\bf v} = f(x) \frac{\partial}{\partial x} + g (x,y) \frac{\partial}{\partial y}$ and ${\bf w} = \eta _x (x,y) \frac{\partial}{\partial x} + \eta _y (x,y) \frac{\partial}{\partial y}$. Then
\begin{eqnarray*}
	[ {\bf v} , {\bf w} ]Ê& = & \left( f \frac{\partial \eta _x  }{\partial x} + g \frac{\partial \eta _x }{\partial y} - \eta_x \frac{\partial f}{\partial x} \right) \frac{\partial}{\partial x} + \\
	&&  \left( f \frac{\partial \eta _y  }{\partial x} + g \frac{\partial \eta _y }{\partial y} - \eta _x \frac{\partial g}{\partial x} - \eta _y \frac{\partial g}{\partial y} \right) \frac{\partial}{\partial y } ,
\end{eqnarray*}
and the condition $[ {\bf v} , {\bf w} ]Ê= 0 $ gives two partial differential equations
\begin{eqnarray*}
	 f \frac{\partial \eta _x  }{\partial x} + g \frac{\partial \eta _x }{\partial y} & = &  \eta_x \frac{\partial f}{\partial x} \\
	 f \frac{\partial \eta _y  }{\partial x} + g \frac{\partial \eta _y }{\partial y}  & = & \eta _x \frac{\partial g}{\partial x} + \eta _y \frac{\partial g}{\partial y}  .
\end{eqnarray*} 
We have the two trivial solutions $\eta _x = f$ , $\eta _y = g$, and $\eta _x = \eta _y = 0$. Since the first equation is fulfilled if $\eta _x = 0$, independent of $\eta _y$, a non-trivial solution can be found if
\begin{eqnarray}
	 f \frac{\partial \eta _y  }{\partial x} + g \frac{\partial \eta _y }{\partial y}  & = & \eta _y \frac{\partial g}{\partial y}  ,
\label{PDE_causal}
\end{eqnarray} 
and $\eta _x =0$. The exact form of the solution is not important since it does not affect the corresponding fiber projection. Formally we write
\begin{eqnarray*}
	{\bf w} ( \pi ) & = & \eta _y (x,y) \frac{\partial \pi (x,y) }{\partial y} = 0 ,
\label{proj_causal}
\end{eqnarray*}
where $\eta _y$ fulfills Eq.~\ref{PDE_causal}. Eq.~\ref{proj_causal} implies
\begin{eqnarray*}
	\pi (x,y) & = & F( x )
\end{eqnarray*}
for an arbitrary  function $F ( \bullet )$. This agrees with our intuition of the fiber projections existing for system. Note that if we try the ansatz $\eta _y = 0$, we get the equations
\begin{eqnarray*}
	 f \frac{\partial \eta _x  }{\partial x} + g \frac{\partial \eta _x }{\partial y} & = &  \eta_x \frac{\partial f}{\partial x} \\
	\eta _x \frac{\partial g}{\partial x} & = & 0 .
\end{eqnarray*} 
The last equation implies either: $\eta _x = 0$, which is a trivial solution; or that $g$ is independent of $x$, which would mean that the original system separates completely and the two degrees of freedom can be studied independently.

As a concrete example, consider the linear case:
\begin{eqnarray*}
	\dot{x} & = & a x + y \\
	\dot{y} & = & a y
\end{eqnarray*}
for some constant $a$. The matrix governing the dynamics is in itself a non-trivial Jordan block and therefore cannot be diagonalized, or reduced. Moreover Schur's lemma, usually referenced in matrix representation theory, see e.g.,~\cite{fuchs}, tells us that the only matrix commuting with the dynamics are the two trivial cases discussed in Section~\ref{gen_linj}, i.e., the dynamics itself and $\zeta \cdot {\bf 1}$. However, non of these symmetries corresponds to the obvious projective map $\pi (x,y) = G(y)$. Using the general condition we can, however, find a nonlinear symmetry:
\begin{eqnarray*}
	{\bf w} & = & x \cdot F \left( \frac{a y - x \log | x | }{a x} \right) \frac{\partial}{\partial x} ,
\end{eqnarray*}
for some arbitrary function $F( \bullet )$, preferably chosen so that ${\bf w}$ is defined everywhere. The result is the projective map $\pi (x , y) = G(y)$, for some function $G ( \bullet )$.
\section{Conclusions and outlook}
In this paper we have not dealt explicitly with topological aspects of projective fiber maps, except for some comments on the map from ${\mathbb R} ^2$ to $S ^1$ in Section~\ref{periodic_orbits}. The general framework for studying these aspects is fiber bundles. The process by which we eliminate subsystems through projective fiber maps is equivalent with the elimination of gauge invariance in modern physics. Gauge theory is one of the richest fields in theoretical physics, so exploring this connection in more detail could clearly lead to interesting results in dynamical systems theory.

Another possible technical complication we have avoided is infinite dimensionality. It is well known that infinite dimensional systems are intrinsically hard hard to analyze. On the other hand, some of the most interesting examples of efficient model reduction, especially though inertial manifolds, stems from infinite dimensional partial differential equations, such as the Navier-Stokes equations.

We have restricted the attention to continuous symmetries. Discrete symmetries, however, are often of central importance, especially in mechanical systems. For example, permutation of identical particles can be used to construct invariant manifolds for $n$-body problems\cite{papenbrock}. Hierarchical dynamics is often very prominent in systems with many identical particles, for example mono atomic gases. One may speculate that this in part is due to the high degrees of symmetries in such systems, primarily through permutations combined with some continuous transformation, such as rotations. From these observations it seems clear that including discrete transformations could be an essential extension of projective fiber map techniques.

It is shown in Prop.~\ref{sym_proj} that the projective fiber maps are directly related to the symmetries of the system. A natural question to ask is whether the structure of the maximal symmetry group of the dynamical system has a natural interpretation in the hierarchical organization of the dynamics. We may speculate that irreducible representations of the symmetry group corresponds to minimal\footnote{Minimal in terms of the number of degrees of freedom that gets eliminated} projective fiber maps. This line of analysis is work in progress.

The main limitation of the current framework is perhaps the restriction that the dynamics on the higher level is required to be deterministic. We know, for example from the success of Langevin-type coarse graining, that allowing stochastic dynamics on the higher levels would greatly extend the applicability of projective fiber maps. Naive attempts at implementing such a scheme, however, result in difficulties. The main problem is to mathematically define strict restrictions on acceptable higher level dynamics. Currently, it seems to the author that we should require the higher level dynamics to be Markovian, i.e., the future time evolution should be determined only by the current state of the system, not by its previous history. This is also in agreement with the conclusions for employing projection operators in non-equilibrium statistical physics\footnote{The noise term in the Langevin equation is usually assumed to be uncorrelated in time.}, see Zwanzig~\cite{zwanzig} for details, as well as the argument by Shalizi and Moore in~\cite{moore}. Recasting this requirement into the differential geometric perspective given here is currently work in progress.

{\bf Acknowledgment}: This work was funded (in part) by PACE (Programmable Artificial Cell Evolution), a European Integrated Project in the EU FP6-IST-FET Complex Systems Initiative, and by EMBIO (Emergent Organisation in Complex Biomolecular Systems), a European Project in the EU FP6 NEST Initiative. The author also acknowledges support from the Swedish Science Foundation (VR). I would  like to thank Steen Rasmussen for initiating this work~\cite{hier}, and also for extensive discussions. I am also grateful to Lina Reichenberg for valuable comments on the manuscript.

\bibliographystyle{unsrt}
\bibliography{articles_dyn_hier.bib}

\appendix

\renewcommand{\theequation}{A-\arabic{equation}}
\setcounter{equation}{1}

\section{Symmetries of ordinary differential equations}
\label{symmetries_ODE}
In this appendix we briefly review the basic theory of Lie group symmetries of ordinary differential equations. For an extensive treatment of this subject, see e.g.,~\cite{olver}. Start by considering a system of algebraic equations
\begin{eqnarray*}
  F _{k} (x) & = & 0 , \;\;\;\;\;\; k = 1, \dots, l,
\end{eqnarray*}
in which $F_1 (x), \dots , F_l (x)$ are smooth real-values functions defined for
$x$ on some manifold $M$. A solution is a point $x \in M$ such that $F_{k} (x) =0$,
$k = 1 , \dots , l$. A symmetry group of the system is a group of transformations
$G$ acting on $M$ such that $G$ transforms solutions to other solutions., i.e.,
if $x$ is a solution and $g$ a group element in $G$ such that $g \cdot x$ is defined,
then $g \cdot x$ is also a solution. The following proposition provides an infinitesimal,
hence more useful, criteria for a Lie group to be a symmetry group:
\begin{proposition}
\label{symm_alg_eq}
Let $G$ be a connected local Lie group of transformations acting on an $m$-dimensional
manifold $M$. Let $F : M \rightarrow {\mathbb R}^l$ define a system of algebraic
equations $F _{k} (x) =  0$, $k = 1, \dots, l$,
of maximal rank. Then $G$ is a symmetry group of the system if and only if
\begin{eqnarray*}
  {\bf w} \left[ F_{k} (x) \right] & = & 0 , \;\; \mbox{whenever} \;\; F_{k} (x) =0
\end{eqnarray*}
for every infinitesimal generator ${\bf w}$ of $G$.
\end{proposition}
{\bf Proof}: Differentiate $F ( \exp ( \epsilon {\bf v} ) x ) = 0$ with respect to $\epsilon$ at $\epsilon = 0$. $\Box$

More generally we say that a function $F : M \rightarrow N$ is $G$-invariant if
$F ( g \cdot x ) = F(x)$ for all $g \in G$ for which $g \cdot x$ is defined. The 
infinitesimal version of this reads:
\begin{proposition}
\label{symm_func}
Let $G$ be a connected local Lie group of transformations acting on an $m$-dimensional
manifold $M$. A smooth function $F : M \rightarrow N$ is $G$-invariant if and only if
\begin{eqnarray*}
  {\bf w} \left[ F \right] & = & 0
\end{eqnarray*}
for all $x$ in $M$ and every infinitesimal generator ${\bf w}$ of $G$.
\end{proposition}
Note that $G$-invariance of a function $F$ is a more strict requirement than 
$G$ being a symmetry group of the system of algebraic equations $F(x) = 0$. In 
fact $F$ being $G$ invariant is equivalent to requiring that $G$ is a symmetry group
of the algebraic equation formed by every level set $\{ F(x) = c \}$, $c \in 
{\mathbb R}^l$. We will need both these formulations later.

The idea behind symmetry analysis of differential equations is essentially equivalent 
to symmetries of algebraic equations, we want to find transformations that map
solutions into new solutions. Ultimately we would like to find criteria of the same type
as Proposition~\ref{symm_alg_eq}. To achieve this, we first need to define a natural 
geometric setting for differential equations. Consider an $n$-th order system of 
differential equations
\begin{eqnarray}
  \dot{x} & = & \xi ( t , x ) ,
\label{diff_eq}
\end{eqnarray}
where $\xi : {\mathbb R}^{n+1} \rightarrow {\mathbb R}^n$ is a smooth function. A solution
to Eq.~\ref{diff_eq} is a graph in the configuration space $\Gamma _{\xi} = \{ 
\left( t , x(t) \right)  : t \in \Omega \} \subset T \times M$, where $\Omega$ if the domain within 
which Eq.~\ref{diff_eq} is defined.

A Lie group transformation $G$ acting on $T \times M$, naturally transforms a solution graph 
according to
\begin{eqnarray}
  g \cdot \Gamma _{\xi} & = & \{ g \cdot ( t, x ) : ( t , x ) \in \Gamma _{\xi} \} .
\label{trans_graph}
\end{eqnarray}
To continue we transform Eq.\ref{diff_eq} into a more geometric form, i.e., an algebraic
relation on some manifold structure. For this en we introduce an simplified
version of a jet space, which in turn is a special case of a fiber space. The first jet
space of the manifold $M$ is defined as Cartesian product manifold
\begin{eqnarray*}
  M ^{(1)} & \equiv & T \times M \times M _1 ,
\end{eqnarray*}
where $M$ is the configuration space and $M _1$ is a space whose coordinates represents
the first derivate of functions with domain in $T$ and range in $M$. Now, since
$t \in T$, $x \in M$, and $\dot{x} \in M_1$, Eq.~\ref{diff_eq} comes an algebraic
equation on $M^{(1)}$, which we denote as:
\begin{eqnarray}
  \Delta _k ( t , x , \dot{x} ) & = & 0 , \;\;\;\;\;\; k = 1, \dots, n .
\label{diff_to_alg}
\end{eqnarray}
The differential equation is now transformed into a geometric (algebraic) form. 
A technical issue remains regarding how a transformation acting on $T \times M$
acts on $M^{(1)}$, i.e., the induced action on $M_1$. As usual the analysis
can be simplified by considering the infinitesimal generators ${\bf w} _i$ of the Lie 
group $G$ rather than the global group transformations from Eq.~\ref{trans_graph}.
Corresponding to each vector field ${\bf w}$ over $T \times M$ we therefore introduce 
a {\em prolongated} vector field ${\bf pr } ^{(1)} {\bf w}$, defined over $M^{(1)}$.
Symmetry group transformations can now be found in terms of prolongated infinitesimal 
generating vector fields:
\begin{proposition}
\label{symm_diff_eq}
Suppose $\Delta _k (t, x , \dot{x} ) = 0$, $k = 1 , \dots , n$, is an $n$-th order system of 
differential equations of maximal rank. Then if $G$ is a local Lie group of transformations 
acting on $M$, and
\begin{eqnarray*}
  {\bf pr} ^{(1)} {\bf w} _i \left[ \Delta _k (t, x , \dot{x} ) \right] & = & 0 ,
\end{eqnarray*}
wherever $\Delta _k (t, x , \dot{x} ) = 0$, $k = 1 , \dots , n$, and for all infinitesimal 
generators ${\bf w} _i$ of $G$, then $G$ is a symmetry group of the system.
\end{proposition}
Deriving an explicit expression of ${\bf pr} ^{(1)} {\bf w}$ as a function of ${\bf w}$
is a somewhat technical matter and beyond the scope of the current presentation
(we refer to Olver~\cite{olver} for details), here we just give the result:
\begin{proposition}
\label{first_prol}
Let 
\begin{eqnarray*}
  {\bf w} & = & \tau ( t , x ) \frac{\partial }{\partial t} +   \sum _{\alpha =1}^n \eta ^{\alpha}
( t , x ) \frac{\partial}{\partial x^{\alpha}}
\end{eqnarray*}
be a vector field defined over
$T \times M$, then the corresponding first prolongated vector field is given by
\begin{eqnarray*}
  {\bf pr} ^{(1)} {\bf w} & = & \tau ( t , x ) \frac{\partial }{\partial t} +   
  \sum _{\alpha =1}^n \eta ^{\alpha} ( t , x ) \frac{\partial}{\partial x^{\alpha}} + \\
 & &   \sum _{\alpha , \beta =1} ^n \left( \dot{x} ^{\beta} \frac{\partial \eta ^{\alpha} (x)}
       {\partial x^{\beta}} \right) \frac{\partial }{\partial \dot{x} ^{\alpha}} ,
\end{eqnarray*}
and defined over $M^{(1)}$.
\end{proposition}
In this paper we only consider systems of autonomous ordinary differential equations. The 
following proposition shows the action of the first prolongation of a generic vector field in
this special case:

\begin{proposition}
\label{prolongation_of_autonomous}
Let ${\bf v}  =  \sum _{\alpha = 1} ^n \xi ^{\alpha} (x) \frac{\partial}{\partial x^{\alpha}}$
be a vector field generating a flow on a manifold $M$. The action of the first prolongation of 
some other vector field ${\bf w}$ on the system of ordinary differential equations 
$\Delta _{k} (t , x , \dot{x} )$, corresponding to the vector field ${\bf v}$, i.e.,
$\dot{x} ^k = \xi ^k (x)$ is given by
\begin{eqnarray*}
  \left. {\bf pr} ^{(1)} {\bf w} \left[ \dot{x} - \xi (x) \right] \right| _x & = & 
  \left. \left[ {\bf v} , {\bf w} \right] \right| _x .
\end{eqnarray*}
\end{proposition}
The following corollary is a direct consequence of Proposition~\ref{prolongation_of_autonomous} 
and~\ref{symm_diff_eq}:

\begin{corollary}
\label{sym_ODE}
Let $G$ be a Lie group, ${\mathfrak g}$ the corresponding Lie algebra, and 
$\{ {\bf w} _i \} _{i=1}^k$ a set of vector fields spanning ${\mathfrak g}$.
A system of autonomous ordinary differential equations with a vector field ${\bf v}$
as infinitesimal generator has $G$ as a symmetry group if and only if
\begin{eqnarray*} 
  \left[ {\bf v} , {\bf w} _i \right] & = & 0 ,
\end{eqnarray*}
$i = 1 , \dots , k$.
\end{corollary}

\end{document}